\documentclass[11pt]{amsart}
\usepackage{geometry}                
\usepackage{graphicx}
\usepackage{amssymb}
\usepackage{amsaddr}
\usepackage{epstopdf}
\usepackage[utf8]{inputenc}
\usepackage{hyperref}
\usepackage{tikz}
\usetikzlibrary{decorations.pathmorphing}
\usepackage{natbib}
\usepackage{algorithm,algpseudocode}
\usepackage{makecell}

\title{On the thermodynamics of prediction under dissipative adaptation}

\author{Kai Ueltzhöffer$^{1,2}$}
\address{$^1$Department of General Psychiatry, Centre of Psychosocial Medicine, University Hospital Heidelberg, Germany} 
\address{$^2$Theoretical Neurobiology Group, Wellcome Centre for Human Neuroimaging, University College London, UK}
\email{kueltzho@gmail.com}
\date{\today}                                           

\begin{document}


\begin{abstract}
On the one hand, the dissipated heat of a thermodynamic work extraction process upper bounds the non-predictive information, which the associated system encodes about its environment. Thus, emergent information processing capabilities can be understood from the perspective of a  pressure towards high thermodynamic efficiency. On the other hand, the second law of thermodynamics plays a crucial role in the emergence of complex, self-organising dissipative structures. Such structures are thermodynamically favoured, because they can dissipate free energy reservoirs, which would not be accessible otherwise. Thereby, they allow a closed system to move from one meta-stable state to another meta-stable state of higher entropy. This paper will argue, that these two views are not contradictory, but that their combination allows to understand the transition from simple self-organising dissipative structures to complex information processing systems. If the efficiency required by a dissipative structure to harvest enough work from the channeled flow of free energy to maintain its own structure is high, there is a drive for this system to be predictive. Still, the existence of this dissipative system is thermodynamically favoured, compared to a situation without any dissipative structure. Due to the emergence of a hierarchy of dissipative systems, which by themselves are non-equilibrium structures that can be dissipated, such a drive develops naturally, as one ascends in this hierarchy further and further away from the initial driving disequilibrium.
\end{abstract}

\maketitle

\section{Introduction}

\subsection{Thermodynamic efficiency implies predictive inference}

The average work $\left< W_\mathrm{diss} \right>$ dissipated by a thermodynamic system, responding to a stochastic driving signal, can be lower bound by the non-predictive information $I_\mathrm{np}$  about the driving signal, which is stored in the state of the system \citep{Still_etal_2012}
$$I_\mathrm{np}=I_\mathrm{mem}-I_\mathrm{pred} \leq \beta \left< W_\mathrm{diss} \right>$$
where $\beta = 1/(k_B T)$, $T$ denotes the temperature of the environment, and $k_B$ the Boltzmann constant. The encoded information $I_\mathrm{mem}=\mathrm{MI}(x_t,s_t)$ is the mutual information between the environmental state $x_t$ and the system state $s_t$ at the same time point. The predictive information $I_\mathrm{pred}=\mathrm{MI}(x_{t+1},s_t)$ is the mutual information between the next environmental state $ x_{t+1}$ and the current system state $s_t$.
Thus, any system which operates with sufficient efficiency will have to encode a compressed, predictive representation of environmental dynamics, to be able to bound the dissipation of excess heat into the environment. This is a powerful argument for the emergence of predictive ”agents”, given the existence of sufficiently complex physical systems, implementing these agents, and  given an environmental drive for such thermodynamic systems to be as efficient as possible. 
While it might seem paradoxical at first, in this work we will argue, that this drive towards efficiency can be directly motivated by the second law of thermodynamics, and the fact, that our universe started in a highly frustrated, meta-stable state of low entropy, following the arguments outlined in the next section.

\subsection{Self-organisation of complex, dissipative structures is directly driven by the second law of thermodynamics}
The emergent self-organisation of dissipative structures in driven non-equilibrium systems has been studied since the early work of Prigogine and colleagues \citep{Nicolis_Prigogine_1977}. Their results were not only applied to self-organising dissipative systems in physics, like convection cells, flames, mechanical cracks, or lightning strikes, but also to biological systems, which also can be conceived as engines or channels to dissipate free energy stored in macroscopic potential gradients \citep{Cottrell_1979, Morowitz_Smith_2007, Branscomb_Russell_2013, Russel_etal_2013}. The core argument is the following: Seemingly unlikely structures are thermodynamically favoured, whenever they open channels for the whole system to transition from one meta-stable state to another meta-stable state of higher entropy, by allowing to dissipate reservoirs of free energy, which would not be accessible otherwise \citep{Jeffery_etal_2019}. Given the fact that our universe started in a highly frustrated state of very low entropy, and that its state space allows for very complex structures, and can features many complex, nested, meta-stable states, the emergence of dissipative structures is a commonly observed phenomenon across scales, from galaxy formation to living cells \citep{Jeffery_etal_2019}.

This argument was given a more quantitative form, building on microscopic fluctuation theorems \citep{Crooks_1999, Jarzynski_1997}, which yield explicit thermodynamic equalities arbitrarily far from equilibrium. Using a coarse-graining method, the following explicit relationship for the relative probability of the forward transition $\pi_\tau$ from one macroscopic state $I$ to two possible outcome states, $II$, $III$, during a finite time $\tau$, was derived \citep{Perunov_etal_2016}:

$$\frac{\pi_\tau(I\rightarrow II)}{\pi_\tau(I\rightarrow III)} = e^{-\beta \Delta E_{II,III}} \frac{\pi^*_\tau(II^*\rightarrow I^*)}{\pi^*_\tau(III^*\rightarrow I^*)}\frac{\left<\exp(-\beta W_\mathrm{diss}\right>_{I\rightarrow III}}{\left<\exp(-\beta W_\mathrm{diss}\right>_{I\rightarrow II}} $$

This generalisation of the Boltzmann distribution contains a standard Boltzmann term as a first factor. The second factor quantifies the kinetic accessibility of the final state, in terms of the reverse transition probabilities $\pi^*_\tau$, owing to the fact that this relation considers transitions over a finite time $\tau$. The third factor quantifies the averaged exponential of the work dissipated along the possible trajectories between the initial and the final states. Thus, given similar energy levels and similar kinetic accessibility, outcome states whose history shows a more reliable dissipation of work are more likely. These theoretical results were recently corroborated by a range of simulation studies, showing spontaneous self-organisation of driven systems into states determined by their dissipative history \citep{Horowitz_England_2017, Kachman_etal_2017}. The ensuing selection process is often called dissipative adaptation.

\subsection{The informational nature of biological systems} 
Although the recent work on dissipative adaptation yields considerate advances in the understanding of the emergence of complex structures in driven thermodynamic systems, one important question remains: What distinguishes living, intelligent systems from other highly stable, emergent, dissipative structures, such as stars, convection cells (including Jupiter’s Great Red Spot), lightning strikes, flames, or mechanical cracks? The answer cannot only lie in the spatial or temporal scales of the processes alone, since the sun implements a relatively simple dissipative system, that extends on both scales far beyond our biosphere.
Recent work hints at a crucial role of information and the astonishing information processing capabilities associated with biological systems \citep{Walker_Davies_2013}, ranging from the long term storage and optimisation of information in the genetic code, to the astonishing information processing capabilities of our central nervous system \citep{Friston_2010}. 

\subsection{An apparent paradox}
The results presented in the first section of the introduction, showing the emergence of information processing capabilities, in terms of a compressed encoding of predictive information, in thermodynamic systems under a pressure to be thermodynamically efficient, might close this explanatory gap \citep{Still_etal_2012}. However, there is an apparent contradiction with the outlined arguments based on the dissipation-driven emergence of complex structures in the origins of life \citep{Jeffery_etal_2019, Morowitz_Smith_2007, Perunov_etal_2016}: How can the second law, which systematically favours outcome states of higher entropy, i.e. histories of higher dissipation, favour the emergence of thermodynamically efficient structures? In the following section, we will try to answer this question.

\section{Results}

\subsection{Outline of the main argument}

The core idea of this argument seems rather simple, but to the best of our knowledge, it has not been stated in this explicit form before: Self-organising dissipative systems have to extract the work required to maintain their own structure (in terms of the associated non-equilibrium steady state) from the potential difference, which is provided by the very free energy reservoir they are dissipating. This required amount work is lower bound by a minimum energetic cost, which only depends on the dissipative system’s non-equilibrium steady state. Now, in situations where the maximum amount of work, which can be extracted from the channelled energy flow, is only marginally larger than the minimum energetic cost required to maintain a dissipative structure, this dissipative structure is still thermodynamically favoured compared to having no dissipative structure at all. We will further argue, that this situation arises naturally in complex systems, given enough time, due to an emergent hierarchy of higher order disequilibria \citep{Branscomb_Russell_2013, Russel_etal_2013} feeding on smaller and smaller potential differences, as shown in Figure \ref{fig1}.

\subsection{Minimum energetic cost to maintain a non-equilibrium steady state}

The minimum energetic cost to maintain a target non-equilibrium state, given by a distribution $p^*$ on the microscopic state space, depends on the distance of this distribution from the equilibrium distribution $p^\mathrm{eq}$, as given by the Kullback-Leibler divergence, and on the equilibrium dynamics of the system. It can be expressed in terms of the minimum entropy production rate $\dot{S}_\mathrm{min}$, which would be observed, if the drive suddenly stopped and the system would start to relax from its target non-equilibrium steady state $p^*$ to its equilibrium steady state $p^\mathrm{eq}$, given its equilibrium dynamics \citep{Horowitz_etal_2017}:

$$ \beta \dot{W}_\mathrm{min} = \dot{S}_\mathrm{min} = \left. \partial^\mathrm{eq}_t D_\mathrm{KL}(p(t)||p^\mathrm{eq})\right|_{p(t)=p^*}$$

Here $\partial^\mathrm{eq}_t$ corresponds to the partial time derivative with respect to the equilibrium dynamics, i.e. as if the system was initialised in a non-equilibrium steady state $p^*$ and then allowed to relax to the equilibrium state $p^\mathrm{eq}$ following the non-driven dynamics.
In general, the required work could also be supplied by an external controller, tasked to maintain a specified non-equilibrium steady state. However, the class of self-organising, auto-catalytic dissipative processes considered here has to harvest this work directly from the dissipated free energy reservoir, via turnstile or bifurcation-like mechanisms \citep{Branscomb_Russell_2013, Cottrell_1979, Nicolis_Prigogine_1977}, as shown in Figure \ref{fig1}B.

\subsection{An emergent hierarchy of disequilibria}

\begin{figure}
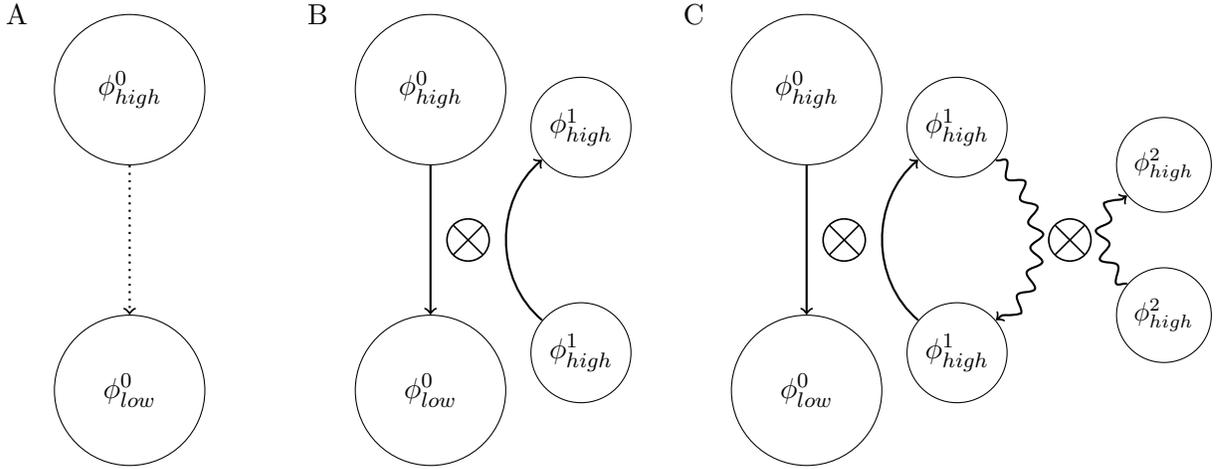


\include{fig1}

\caption{\textbf{A} A reservoir of free energy, stored in a potential difference between $\phi_\mathrm{high}^0$ and $\phi_\mathrm{low}^0$ . In a meta-stable state, there is no simple way to realise flows to dissipate this disequilibrium, as indicated by the dotted arrow. \textbf{B} Given sufficient time and complexity of the state space, dissipative structures can arise, which allow a dissipative flow along the potential gradient of the driving disequilibrium. However, these dissipative structures correspond to non-equilibrium steady states themselves, i.e. they create a new, higher order disequilibrium, with an associated potential difference between $\phi_\mathrm{high}^1$ and $\phi_\mathrm{low}^1$. Crucially, in the self-organizing class of systems considered here, the dissipative structures have to harvest the work required to maintain their non-equilibrium steady state directly from the conducted energy flow, via bifurcation or turnstile-like coupling mechanisms, symbolized by $\otimes$. The corresponding energy flows are symbolised by the straight arrows. C. Given even more time and sufficient complexity of the environment, new dissipative structures can form, dissipating the higher order disequilibrium created by the initial dissipative structure. Another layer of coupled energy flows, symbolized by the wiggly lines, is required to maintain a new level of dissipative structures, with new associated potentials $\phi_\mathrm{high}^2$ and $\phi_\mathrm{low}^2$. Given enough time and complexity of the environment, this process can be iterated, until the minimum amount of work required to maintain a new dissipative structure exceeds the work that could be harvested from the potential difference provided by the final level of existing dissipative structures.}
\label{fig1}

\end{figure}

Free energy reservoirs correspond to disequilibria, i.e. meta-stable states of lower-than-maximum entropy, often due to macroscopic structures, that want to relax to meta-stable states of higher entropy (and finally to equilibrium). Dissipative systems themselves, during the dissipation of their driving disequilibrium, create new, higher order disequilibria in terms of their own, macroscopic structures, which exist in a steady state far from equilibrium. Thus, given a sufficient amount of time and sufficient complexity of a system’s state space, new dissipative systems can form, which dissipate these higher order disequilibria \citep{Branscomb_Russell_2013, Russel_etal_2013}, as shown in Figure \ref{fig1}A-C. An example of such a hierarchy might start with our sun, which dissipates nuclear energy stored in hydrogen atoms, by fusing them to helium nuclei. From the perspective of this process, the outgoing ultra violet radiation corresponds to a significant increase in entropy. However, due to the resulting, new disequilibrium between the high-energy radiation from our sun, and the relatively cool night sky, the biosphere on earth is able to harvest some of the corresponding free energy by turning low wavelength solar photons, into high wavelength infrared photons, thereby increasing their entropy by another factor of about 20.

\subsection{Increasing drive towards efficiency for higher order disequilibria}

The available potential differences associated with higher order disequilibria get smaller and smaller, the further one moves away from the initial, fundamental driving disequilibria. However, the minimum work required to maintain a non-equilibrium steady state, corresponding to a dissipative structure at higher levels, does not necessarily decrease. Instead, it is even conceivable that, to harvest smaller and smaller potential differences, less and less likely configurations in state space are necessary, which might lead to an increase in the Kullback-Leibler-divergence between the corresponding non-equilibrium steady states and the equilibrium state of the dissipative system, $D_\mathrm{KL}(p^*)||p^\mathrm{eq})$, thereby increasing the minimum work required to maintain such a dissipative structure. Thus, the drive for dissipative structures to efficiency is likely to increase, as one ascends in the emergent hierarchy of disequilibria further away from the initial driving disequilibria. Thereby, the dissipative structures emerging on the highest levels of this hierarchy of disequilibria will have to be very efficient, to harvest enough energy from the potentials provided by lower level dissipative structures to sustain their own, non-equilibrium steady states. But crucially, over long enough time scales, such structures will still be thermodynamically favoured compared to no dissipative structures at this level.

\section{Discussion}

Our argument reconciles the seemingly contradictory perspectives of spontaneous structure formation, driven by the second law of thermodynamics, i.e. the dissipation of free energy reservoirs and the associated increase in entropy, and the fact that we see an abundance of highly efficient, thermodynamic machinery in living systems, characterised by information processing capabilities hardly found in the non-living nature. It is based on two observations, namely 1.) that highly efficient dissipative systems are still favoured by the second law, compared to no dissipative systems, and 2.) that the emergent hierarchy of dissipative systems, given enough time and complexity of the state space, will lead to an increasing pressure to get more and more efficient towards the highest levels of the hierarchy, where the initial free energy reservoir is already substantially degraded. 

\subsection{Connection to a hierarchy of phase transitions from a chemical to an information driven regime}

The described hierarchy of dissipative systems has to form in a sequence of fluctuation driven, symmetry breaking phase transitions, starting from the initial, driving disequilibria. Therefore, the higher levels, experiencing an increasing pressure towards thermodynamic efficiency, emerge after the lower ones. This might also explain the sequence of emergence of increasingly information-driven biological systems, such as the genetic code, or fast central nervous systems, in the nested hierarchy of biological phase-transitions and associated structure forming processes, which putatively started with a relatively simple core metabolism \citep{Smith_Morowitz_2016}.

\subsection{Directions for future work}
Our argument predicts the emergence of dissipative structures, showing predictive, agent-like behaviour in (simulation) experiments, implementing the following set of constraints: 

\begin{itemize}
\item A free energy reservoir, which cannot be easily dissipated, and which features some non-trivial temporal dynamics, whose knowledge make it easier to access.
\item A sufficiently complex state space, allowing for a priori very unlikely non-equilibrium steady states, which form channels for the free energy reservoir to be dissipated, and which can feature complex temporal dynamics. 
\item A value of the maximum amount of work, which can be extracted from the free energy reservoir by a dissipative structure, only marginally larger than the minimum work required to maintain such a dissipative structure, to create a pressure on the efficiency of the dissipative structures. 
\end{itemize}
Thereby, only dissipative structures, which are efficient enough to extract the work required to maintain their own structure from the conducted energy flow can exist over extended periods of time. These constraints might model the pressure towards efficiency at the highest levels of a dissipative hierarchy, which might be infeasible to simulate in its entirety, due to limited compute time and resources.

\section{Acknowledgements} I'm indebted to Lancelot Da Costa, Conor Heins, and Annina Luck for fruitful, inspiring discussions, to Sabine Herpertz and Karl Friston for hosting me in their groups at Heidelberg and London, and to the DAAD PRIME initiative for funding, flexibility and support during these uncertain times.

\bibliography{references}

\begin{thebibliography}{}

\bibitem[Branscomb and Russell, 2013]{Branscomb_Russell_2013}
Branscomb, E. and Russell, M.~J. (2013).
\newblock Turnstiles and bifurcators: the disequilibrium converting engines
  that put metabolism on the road.
\newblock {\em Biochim Biophys Acta}, 1827(2):62--78.

\bibitem[Cottrell, 1979]{Cottrell_1979}
Cottrell, A. (1979).
\newblock The natural philosophy of engines.
\newblock {\em Contemporary Physics}, 20(1):1--10.

\bibitem[Crooks, 1999]{Crooks_1999}
Crooks, G.~E. (1999).
\newblock Entropy production fluctuation theorem and the nonequilibrium work
  relation for free energy differences.
\newblock {\em Phys. Rev. E}, 60:2721--2726.

\bibitem[Friston, 2010]{Friston_2010}
Friston, K. (2010).
\newblock The free-energy principle: a unified brain theory?
\newblock {\em Nature Reviews Neuroscience}, 11(2):127--138.

\bibitem[Horowitz and England, 2017]{Horowitz_England_2017}
Horowitz, J.~M. and England, J.~L. (2017).
\newblock Spontaneous fine-tuning to environment in many-species chemical
  reaction networks.
\newblock {\em Proceedings of the National Academy of Sciences},
  114(29):7565--7570.

\bibitem[Horowitz et~al., 2017]{Horowitz_etal_2017}
Horowitz, J.~M., Zhou, K., and England, J.~L. (2017).
\newblock Minimum energetic cost to maintain a target nonequilibrium state.
\newblock {\em Phys. Rev. E}, 95:042102.

\bibitem[Jarzynski, 1997]{Jarzynski_1997}
Jarzynski, C. (1997).
\newblock Nonequilibrium equality for free energy differences.
\newblock {\em Phys. Rev. Lett.}, 78:2690--2693.

\bibitem[Jeffery et~al., 2019]{Jeffery_etal_2019}
Jeffery, K., Pollack, R., and Rovelli, C. (2019).
\newblock On the statistical mechanics of life: Schr{\"o}dinger revisited.
\newblock {\em Entropy}, 21(12).

\bibitem[Kachman et~al., 2017]{Kachman_etal_2017}
Kachman, T., Owen, J.~A., and England, J.~L. (2017).
\newblock Self-organized resonance during search of a diverse chemical space.
\newblock {\em Phys. Rev. Lett.}, 119:038001.

\bibitem[Morowitz and Smith, 2007]{Morowitz_Smith_2007}
Morowitz, H. and Smith, E. (2007).
\newblock Energy flow and the organization of life.
\newblock {\em Complexity}, 13:51--59.

\bibitem[Nicolis and Prigogine, 1977]{Nicolis_Prigogine_1977}
Nicolis, G. and Prigogine, I. (1977).
\newblock {\em Self-organization in nonequilibrium systems : from dissipative
  structures to order through fluctuations / G. Nicolis, I. Prigogine}.
\newblock Wiley, New York.

\bibitem[Perunov et~al., 2016]{Perunov_etal_2016}
Perunov, N., Marsland, R.~A., and England, J.~L. (2016).
\newblock Statistical physics of adaptation.
\newblock {\em Phys. Rev. X}, 6:021036.

\bibitem[Russell et~al., 2013]{Russel_etal_2013}
Russell, M.~J., Nitschke, W., and Branscomb, E. (2013).
\newblock The inevitable journey to being.
\newblock {\em Philosophical Transactions of the Royal Society B: Biological
  Sciences}, 368(1622):20120254.

\bibitem[Smith and Morowitz, 2016]{Smith_Morowitz_2016}
Smith, E. and Morowitz, H.~J. (2016).
\newblock {\em The Origin and Nature of Life on Earth: The Emergence of the
  Fourth Geosphere}.
\newblock Cambridge University Press.

\bibitem[Still et~al., 2012]{Still_etal_2012}
Still, S., Sivak, D.~A., Bell, A.~J., and Crooks, G.~E. (2012).
\newblock Thermodynamics of prediction.
\newblock {\em Phys Rev Lett}, 109(12):120604.

\bibitem[Walker and Davies, 2013]{Walker_Davies_2013}
Walker, S.~I. and Davies, P. C.~W. (2013).
\newblock The algorithmic origins of life.
\newblock {\em Journal of The Royal Society Interface}, 10(79):20120869.

\end{thebibliography}
\bibliographystyle{apalike}

\end{document}